\documentclass[aip,amsmath,amssymb,a4paper,reprint]{revtex4-1}
\usepackage{natbib}
\usepackage[pdftex]{graphicx} 
\usepackage{epstopdf} 
\usepackage{dcolumn} 
\usepackage{bm} 
\usepackage{color}
\usepackage{xspace} 
\usepackage{textcomp} 
\usepackage[version=3]{mhchem} 
\usepackage{lmodern} 
\usepackage{microtype} 
\usepackage{hyperref} 
\renewcommand{\mu}{\text \textmu} 
\usepackage{upgreek} 

\begin{document}

\title{Growth, characterization, and transport properties of ternary (Bi$_{1-x}$Sb$_x$)$_2$Te$_3$ topological insulator layers}

\affiliation{Peter Gr\"unberg Institute (PGI-9) and JARA-Fundamentals of Future Information
	Technology, Forschungszentrum J\"ulich GmbH, 52425 J\"ulich, Germany}
\affiliation{2nd Institute of Physics and JARA-Fundamentals of Future Information
	Technology, RWTH Aachen University, 52074 Aachen, Germany}
\affiliation{Peter Gr\"unberg Institute (PGI-6) and JARA-Fundamentals of Future Information
	Technology, Forschungszentrum J\"ulich GmbH, 52425 J\"ulich, Germany}
\affiliation{Helmholtz Virtual Institute for Topological Insulators (VITI), Forschungszentrum J\"ulich, 52425 J\"ulich, Germany}
\affiliation{Helmholtz Virtual Institute for Topological Insulators (VITI), RWTH Aachen, 52074 Aachen, Germany}

\author{C. Weyrich,$^{1,4}$ M. Dr\"ogeler,$^{2,5}$ J. Kampmeier,$^{1,4}$ M. Eschbach,$^{3,4}$ 
	G. Mussler,$^{1,4}$ T. Merzenich,$^{1,4}$ T. Stoica,$^{1}$ I. E. Batov,$^{1,4}$ 
	J. Schubert,$^{1}$ L. Plucinski,$^{3,4}$ B. Beschoten,$^{2,5}$ C. M. Schneider,$^{3}$ 
	C. Stampfer,$^{1,2,5}$ D. Gr\"utzmacher,$^{1,4}$}
\author{Th. Sch\"apers$^{1,4,}$}
\email{th.schaepers@fz-juelich.de} 


\begin{abstract}
Ternary (Bi$_{1-x}$Sb$_x$)$_2$Te$_3$ films with an Sb content between 0 and 100\% were deposited on a Si(111) substrate by means of molecular beam epitaxy. X-ray diffraction measurements confirm single crystal growth in all cases. The Sb content is determined by X-ray photoelectron spectroscopy. Consistent values of the Sb content are obtained from Raman spectroscopy. Scanning Raman spectroscopy reveals that the (Bi$_{1-x}$Sb$_x$)$_2$Te$_3$ layers with an intermediate Sb content show spatial composition inhomogeneities. The observed spectra broadening in angular-resolved photoemission spectroscopy (ARPES) is also attributed to this phenomena. Upon increasing the Sb content from $x=0$ to 1 the ARPES measurements show a shift of the Fermi level from the conduction band to the valence band. This shift is also confirmed by corresponding magnetotransport measurements where the conductance changes from $n$- to $p$-type. In this transition region, an increase of the resistivity is found, indicating a location of the Fermi level within the band gap region. More detailed measurements in the transition region reveals that the transport takes place in two independent channels. By means of a gate electrode the transport can be changed from $n$- to $p$-type, thus allowing a tuning of the Fermi level within the topologically protected surface states.   
\end{abstract}

\maketitle

\section{Introduction}

The discovery of the existence of topological insulators (TI) has led to increased scientific interest in the study of chalcogenide materials that are found to belong to this class of materials. \citep{Chen09,Hsieh09} Three-dimensional topological insulators are materials with a finite energy bandgap (about $300-400$\,meV) in bulk and zero gap surface states when in contact to a regular insulator. These states are topologically protected due to strong spin-orbit interactions and time reversal invariance.\citep{Culcer12,Hasan11,Qi11,Hasan10} However, the demonstration of TI surface behavior of materials is challenging for both theory and experiment. Theoretical computations have shown a TI character for Bi$_2$Te$_3$, Bi$_2$Se$_3$ and Sb$_2$Te$_3$.\citep{Zhang09} The electronic properties of surface states of TI materials have been studied widely using angle-resolved photoemission spectroscopy (ARPES). \citep{Chen09,Hsieh09,Hsieh09b,Plucinski11,Zhang09,Takagaki12} As in the case of ARPES, surface magneto-electric properties measured by scanning tunneling microscopy (STM) have shown less sensitivity of the protected surface states to intrinsic defects in Sb$_2$Te$_3$ thin films.\citep{Jiang12} In the same article, the thickness limit for three-dimemensional TI was evaluated to four quintuple layers ($\sim$ 4\,nm). For thinner layers the protected Dirac point vanishes because the surface states of opposing sides overlap and a gap opens up due to hybridization.

The robustness of the TI materials to elastic scattering, or non-magnetic impurities, and surface imperfection was the subject of a few investigations,\citep{Plucinski11,Zhang11,Takagaki12,Wray11,Butch10} as it is important for future applications in electronics. Surface electrical transport investigations were hampered by the high density of bulk carriers usually obtained due to intrinsic defects observed in these small gap materials.\citep{Lostak1989,Scanlon2012} Therefore, the effort has concentrated on controlling the sample fabrication and/or carrier compensation doping, or alloying of intrinsic chalcogenide materials, in order to place the Fermi level in the energy bandgap, closer to the Dirac zero gap point of the surface energy spectra.

Magnetoresistive effects, as for example Shubnikov--de Haas oscillations have been measured from surface states of selected cleaved Bi$_2$Te$_3$ crystals with non-metallic behavior \citep{Qu10} or exfoliated BiSbTeSe$_2$ flakes \citep{Xu14}, demonstrating the importance of reducing the bulk carrier concentration for electrical studies of TI features. It was shown that the conductance of Bi$_2$Te$_3$ can be controlled by Sn doping.\citep{Chen09}  Furthermore, the conductance of Bi$_2$Se$_3$ can be tuned from $n$- to $p$-type by doping with Ca.\citep{Hsieh09b} However, the compensation of intrinsic carriers by doping may induce strong perturbation of the surface electronic transport, especially for thin films, in spite of the expected topological protection.\citep{Taskin12} Therefore, in order to reduce the bulk conductance, a precise control of the growth conditions\citep{Jiang12,Wang11,Lee12} or different growth methods and appropriate substrates are desirable.\citep{Jiang12b,Kong10,Liu11,Zeng2013} Another possibility is alloying of similar TIs as for example Bi$_2$Te$_3$ and Sb$_2$Te$_3$.\citep{Zhang11,Ren10,Kong11,He12,Shimizu12} Usually, Bi$_2$Te$_3$ films show $n$-type metallic behavior, while Sb$_2$Te$_3$ has $p$-type conductivity as a result of unintentional defect creation. In both materials the formation of Te-vacancies and, more importantly, Te$_\mathrm{Bi}$- or Sb$_\mathrm{Te}$-antisite defects are considered to be responsible for the intrinsic background doping. Theoretical and experimental studies have shown that the TI surface properties of the solid solution is preserved for the entire composition range of (Bi$_{1-x}$Sb$_x$)$_2$Te$_3$ alloy.\citep{Kong11} The experiments were performed on nano-plates grown using evaporation of Sb$_2$Te$_3$ and Bi$_2$Te$_3$ as precursors in a tube furnace. The transition from $n$-type to $p$-type was found by ARPES and electrical measurements to occur at a composition of about 50\%.\citep{Kong11} Studies on molecular beam epitaxy (MBE) grown (Bi$_{1-x}$Sb$_x$)$_2$Te$_3$ films on sapphire have shown the transition from $n$-type to $p$-type to occur at about 94\% Sb.\citep{Zhang11} In MBE-grown (Bi$_{1-x}$Sb$_x$)$_2$Te$_3$ films on SrTiO$_3$ substrates a minimum carrier density  was found for 50\% Sb by tuning the carrier density with the help of a gate voltage, while the change from $n$-type to $p$-type occurs in the range of 35-45\% Sb at zero gate voltage.\citep{He12}  

We investigated a set of (Bi$_{1-x}$Sb$_x$)$_2$Te$_3$ films grown by MBE on Si(111), with an Sb content $x$ ranging from 0 to 100\%. The layers were first characterized by X-ray diffraction (XRD), ARPES and scanning Raman spectroscopy. Here, detailed  information on structural and electronic properties as a function of composition were gained. Furthermore, magnetotransport measurements were conducted, in order to extract both the carrier concentration and the dominant type of charge carriers, i.e. $p$- or $n$-type. These results were linked to ARPES measurements.
The systematic measurements on samples with different Sb contents allowed us to specify the Sb content, where the compensation of $n$- and $p$-type doping resulted in a suppression of the bulk conductance contribution. Thus, the transport is dominated by contribution of the surface and interface layers. These conclusions are supported by additional weak antilocalization measurements. 

\section{Growth and Experimental details}
(Bi$_{1-x}$Sb$_x$)$_2$Te$_3$ films are deposited on a RCA cleaned Si(111) layer of silicon-on-insulator (SOI) substrates, taking care to remove the native oxide layer before deposition as described in Ref. [\citenum{Krumrain11}] for Bi$_2$Te$_3$ films. Here, antimony was incorporated in parallel by applying an Sb flux additionally to those of Bi and Te. The deposition duration and temperature, as well as the Te flux were constant for all deposited samples, while the ratio of Sb and Bi fluxes was varied. The composition of the (Bi$_{1-x}$Sb$_x$)$_2$Te$_3$ was determined by X-ray photoelectron spectroscopy (XPS). The corresponding values are given in Tab.~\ref{tab:table-1}. The thicknesses $d$ of the films evaluated from X-ray reflectivity measurements vary between 13 and 40\,nm, while the roughness $\delta d$ extracted from atomic force microscopy was in the range from 1.2 to 3.5\,nm (see Tab.~\ref{tab:table-1}). The X-ray reflectivity measurements revealed a good uniformity of the film thickness. The growth rate was varied in the range from 4 to 13\,nm/h.  
\begin{table}
	\caption{Sample parameters: sample, thickness $d$, roughness $\delta d$ and Sb content $x$.}
	\label{tab:table-1}
	\begin{tabular*}{1.0\columnwidth}{@{\extracolsep{\fill}}cccc}
		\hline \hline \\[-7pt]
		Sample	& $d$\,(nm)	& $\delta d$\,(nm)	& Sb content $x$	\\ \\[-8pt]
		\hline \\[-8pt]
		A1		& 13.2		& 1.5				& 0\%				\\
		A2		& 17.5		& 1.5				& 30\% 				\\
		A3		& 22.5		& 1.5				& 49\%				\\
		A4		& 35.0		& 2.8				& 63\%				\\
		A5		& 32.0		& 3.0				& 82\% 				\\
		A6		& 40.0		& 3.5				& 93\% 				\\
		A7		& 27.0		& 2.5				& 95\% 				\\
		A8		& 29.8		& 2.3				& 100\% 			\\ \\[-8pt]
		\hline \\[-8pt]
		B1		& 16.5		& 1.5				& 39\% 				\\
		B2		& 24.0		& 1.4				& 42\% 				\\
		B3		& 27.8		& 1.2				& 43\% 				\\
		B4		& 23.0		& 3.0				& 45\% 				\\ \\[-8pt]
		\hline \hline
	\end{tabular*}
\end{table}

High-resolution ARPES was performed on cleaned surfaces of the ternary thin films. To this end a, single annealing step of 250$^\circ$C for about $5-10$\,min was sufficient to desorb surface contaminations and to provide decent photoemission signals. All high-resolution ARPES spectra presented here were obtained in a laboratory-based system using a monochromatized microwave-driven Xe discharge source producing photons of $h \nu = 8.44$\,eV energy and a beam spot size of about 1\,mm as described in detail by Suga \textit{et al.}\citep{Suga10} The system was recently upgraded with a MBS A1 spectrometer which was set to 20\,meV energy resolution while the samples were cooled to $15-20$\,K.

The Raman measurements were carried out using a WITec alpha300 R confocal Raman microscope. All spectra were taken at room temperature under ambient condition. A green laser with a wavelength of 532\,nm was focused on the sample via a 50$\times$ objective. The resulting spot size was around 400\,nm. The spectra were recorded using a grating with 1800\,lines/mm and a resolution of 0.8\,cm$^{-1}$. Since the investigated films have a low heat conductance and a low melting point a laser power of 200\,$\mu$W was chosen, in order to prevent thermal damages on the samples.

The crystal structure of (Bi$_{1-x}$Sb$_x$)$_2$Te$_3$ consists of stacked quintuple layers that are internally bound by covalent bonds but exhibit only a relatively weak van der Waals interaction between each other. The unit cell is made up of 5 atoms which leads to 15 lattice dynamical modes where three are acoustic and 12 are optical. Four of these 12 optical modes are Raman active, which are two E$_g$ and two A$_{1g}$ modes. In this study we focus on the E$_g^2$ and A$_{1g}^2$ modes as the E$_g^1$ and A$_{1g}^1$ modes have energies below 100\,cm$^{-1}$ which are not accessible in our setup due to the edge filter used to attenuate directly back scattered light. The E$_g^2$ and A$_{1g}^2$ modes are expected to be in the range of 100 to 170\,cm$^{-1}$ and will be discussed in the corresponding section.\citep{Chis12}

The magnetotransport properties of the first series of (Bi$_{1-x}$Sb$_x$)$_2$Te$_3$ layers with an Sb content $x$ varying between 0 and 100\% (samples A1 to A8) were determined using the van der Pauw method. For the Ohmic contacts at each corner of the square-shaped samples a Ti/Au layer was used. The van der Pauw method was chosen because of the simple sample fabrication allowing a quick straightforward assessment of the carrier type and resistivity. After narrowing the range of Sb content close to the transition between $p$- and $n$-type, a new series of layers was grown (samples B1 to B4) and Hall bar structures were defined by optical lithography and Ar$^+$ dry etching. The Hall bar mesa structures had a width of $20-60$\,$\mu$m. Similarly to the van der Pauw samples a Ti/Au layer system was used for the Ohmic contacts. The voltage probes were separated by 150\,$\mu$m. For the sample with an Sb content of 43\% we fabricated a gate electrode covering the complete Hall bar. As a high-$k$ gate dielectric an amorphous 100\,nm thick LaLuO$_3$ layer grown by pulsed laser deposition was used.\citep{Lopes2007}    

The magnetotransport measurements were carried out in a variable temperature insert at temperatures down to 1.8\,K. The magnetic field of up to 13.5\,T was oriented perpendicularly to the sample surface. All measurements were performed in a four-terminal configuration using standard lock-in technique with an AC current of up to $I=25$\,$\mu$A. 

\section{Characterization}

\subsection{XRD}

\begin{figure}[htb]
\includegraphics[width=0.95\columnwidth]{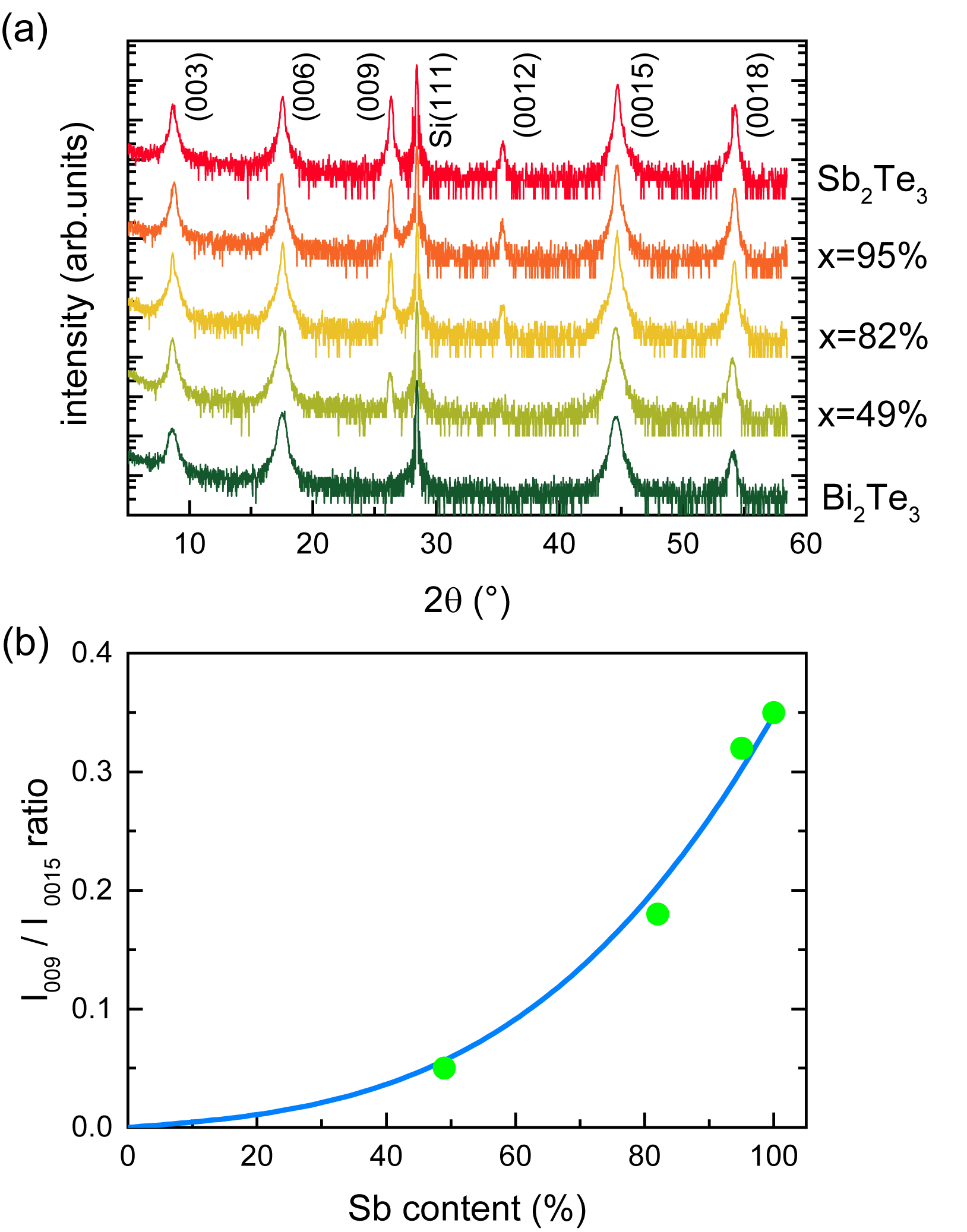}
\caption{a) Symmetric $2\theta /\theta$ scans of (Bi$_{1-x}$Sb$_x$)$_2$Te$_3$ samples with different Sb concentrations. b) Intensity ratio $I_{009}/I_{0015}$ with respect to the Sb content, including a nonlinear fit of its dependence.}
\label{Fig1-XRD}
\end{figure}

To determine the crystal structure, X-ray diffraction was used. Figure \ref{Fig1-XRD}(a) shows symmetric $2\theta /\theta$ scans of (Bi$_{1-x}$Sb$_x$)$_2$Te$_3$ samples with different Sb content. Several peaks are seen which are all collinear with the (001) direction, indicating single-crystal growth with the c-axis in growth direction. The XRD curves of the (Bi$_{1-x}$Sb$_x$)$_2$Te$_3$ alloy depicted in Fig.\ref{Fig1-XRD}(a) show a negligible shift of peak positions as a function of the Sb content. As the lattice constants of Bi$_2$Te$_3$ and Sb$_2$Te$_3$ are almost identical ($c =3.049$\,nm and 3.046\,nm, $a =0.438$\,nm and 0.426\,nm for Bi$_2$Te$_3$ and Sb$_2$Te$_3$, respectively)\citep{Madelung00} the peak positions cannot be used to determine th Sb concentration. However, the diffraction lines corresponding to the (0\,0\,9) and (0\,0\,12) planes are forbidden in Bi$_2$Te$_3$ and not in Sb$_2$Te$_3$, due to the diffraction form factors. Hence, from the $I_{009}/I_{0015}$ intensity ratio, it is feasible to determine the Sb content by means of XRD. Figure \ref{Fig1-XRD}(b) depicts the $I_{009}/I_{0015}$ intensity ratio for different Sb contents. The Sb contents were independently measured by means of X-ray photoelectron spectroscopy (data not shown). A strongly non-linear dependence is seen, and the polynomial fit yields the dependence $0.042\, x + 0.306\, x^3$.

\subsection{ARPES}

\begin{figure*}[htb]
	\includegraphics[width=0.8\textwidth]{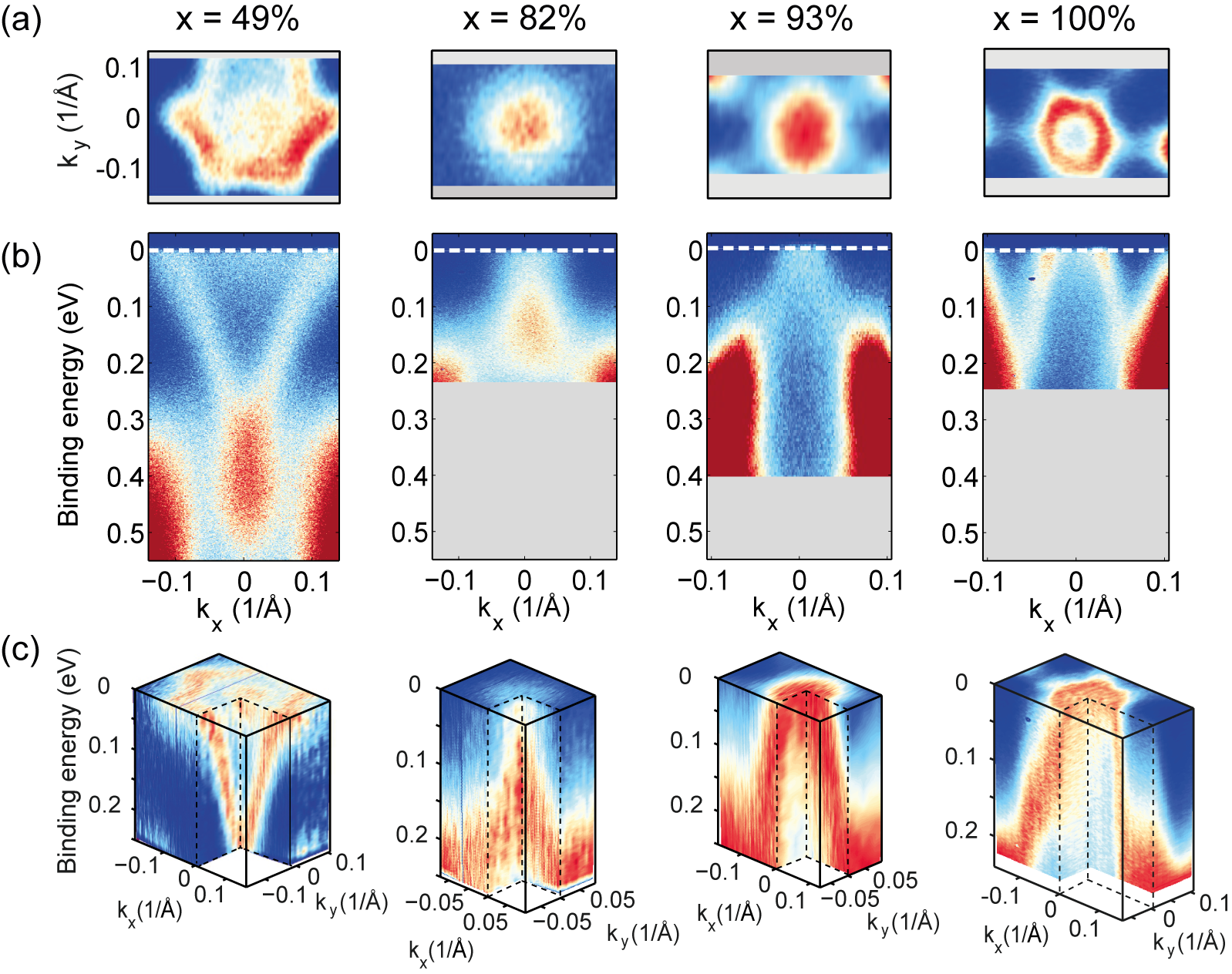}
	\caption{High resolution ARPES spectra obtained by a laboratory based Xe light source ($h\nu = 8.44$\,eV) for x=49\%, 82\%, and 93\% ternary films and a pure Sb$_2$Te$_3$ film, respectively. (a) Fermi surface maps around $\Gamma$ with slightly different cut directions. The hexagonal shape of the Dirac cone above and underneath the Dirac point is clearly visible. (b) Binding energy dispersion spectra at $k_y =0$\,A$^{-1}$ vs $k_x$ at normal emission. The dashed line corresponds to the Fermi energy. (c) Three-dimensional plots of the constant energy surfaces in the Dirac cone region. Note, that the pure Sb$_2$Te$_3$ film does not belong to the above mentioned set of epitaxial layers, but serves as a reference sample.}
	\label{Fig2-ARPES}
\end{figure*}

Figure~\ref{Fig2-ARPES} shows the surface electronic structure of three of the investigated ternary alloys (x=49\%, 82\%, and 93\%) as well as a spectrum from a binary Sb$_ 2$Te$_3$ film (x=100\%), respectively. In each of the spectra depicted in Figs.~\ref{Fig2-ARPES}(b) and (c) the Dirac cone-like topological surface states can unambiguously be detected. The fact that all of the spectral features are rather unsharp and smeared cannot be attributed to the lacking energy or angular resolution but has to be connected to surface quality and more importantly homogeneity of the samples as the signal is averaged over the relatively large spot size of 1\,mm. However, the presented spectra show a clear shift of the electronic structure. In the case of an Sb content of 49\% the electronic structure reveals an $n$-type character with the Fermi level close to the conduction band which is similar to the well known characteristics of pure Bi$_2$Te$_3$, as shown by Plucinski \textit{et al.}\citep{Plucinski11} Further, the entire upper Dirac cone is visible whereas the lower part seems to be buried in the valence band. This is followed by an intermediate state in the 82\% and 93\% Sb content samples where the Dirac point seems to be close to or smeared out around the Fermi level. And finally, pure Sb$_2$Te$_3$ films exhibit a $p$-type character where only the lower Dirac cone is occupied. Furthermore, the Fermi surface maps shown in  Fig.~\ref{Fig2-ARPES}(a) reveal that for the 93\% and 100\% Sb content samples the Fermi level clearly truncates the valence band, thus $p$-type transport is expected. To investigate the stoichiometry around x=93\% in more detail, additional studies where performed on samples that were transferred $in$-$situ$ from the MBE to the ARPES system. The analysis done in Ref. [\citenum{Kellner}] reveals, that for x=94\% the Fermi energy lies within the gap without truncating any bulk bands, shifted only by $\sim5$\,meV from the Dirac point of the surface states. This finding is in good agreement with the results shown in this work.

\subsection{Raman}

\begin{figure*}[htb]
	\includegraphics[width=0.8\textwidth]{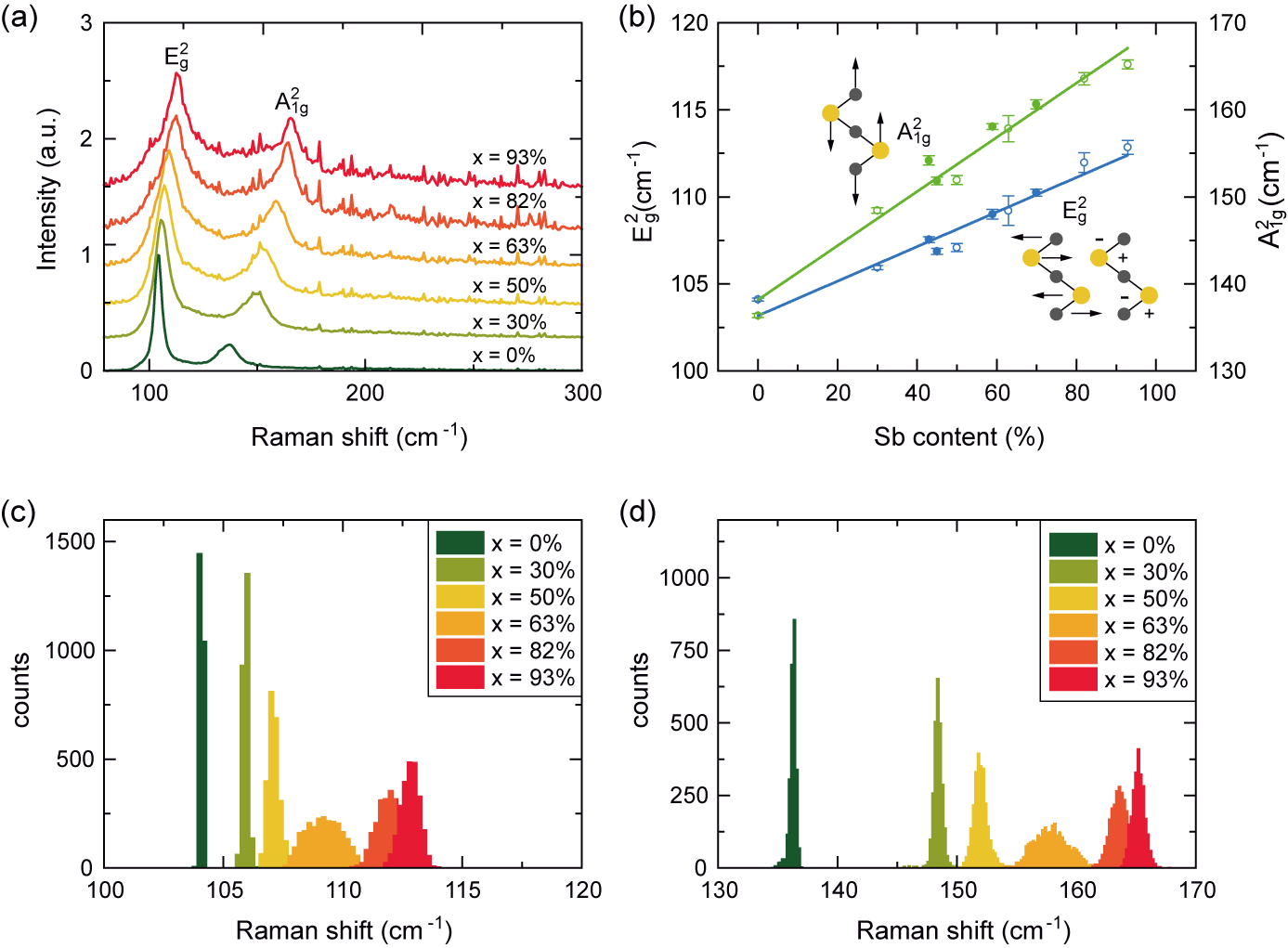}
	\caption{(a) Average spectra for different Sb contents x. All spectra are normalized by the Intensity of the E$_g^2$ mode. (b) Average Raman shift of the E$_g^2$ and the A$_{1g}^2$ mode for different Sb contents $x$. The error bar stems from the width of the distribution of the measured Raman shifts within an area of $20 \times 20\,\mu$m$^2$. The solid symbols belong to the additional samples with optimized growth parameters. (c) and (d) show the corresponding distribution for the E$_g^2$ and the A$_{1g}^2$ mode.}
	\label{Fig3-Raman}
\end{figure*}

Micro Raman data were obtained on (Bi$_{1-x}$Sb$_x$)$_2$Te$_3$ films with varying Sb content. To compare films with different Sb content we recorded a map of $50 \times 50$ spectra within an area of $20 \times 20\,\mu$m$^2$ for every film to give consideration to possible local variations. To first compare the Raman spectra for different Sb contents, the average over all local spectra for each film was calculated. The results are shown in Fig. \ref{Fig3-Raman}(a). These spectra are normalized by the height of the E$_g^2$ peak. It can be observed that both, the E$_g^2$ and the A$_{1g}^2$ peak, shift towards larger wavenumbers for increasing Sb content. To analyze this shift in more detail the E$_g^2$ and the A$_{1g}^2$ peaks of each spectrum are fitted by two Lorentzian peaks and the positions of the peaks are collected in a histogram for each sample which are shown in Figs.~\ref{Fig3-Raman}(c) and (d). These histograms are fitted by Gaussian functions to get both the average peak position and the variation which are summarized in Fig.~\ref{Fig3-Raman}(b). An almost linear relation is observed between the positions of both peaks and the Sb content determined by XPS. This shift can be explained by the different masses Bi and Sb.\citep{Richter77} Since Sb is lighter than Bi, this leads to higher frequencies and therefore larger Raman shifts with increasing Sb content. Thus, complementary to XPS measurements, it is possible to use Raman spectroscopy to estimate the Sb content in a non-destructive way.

\begin{figure*}[htb]
	\includegraphics[width=0.75\textwidth]{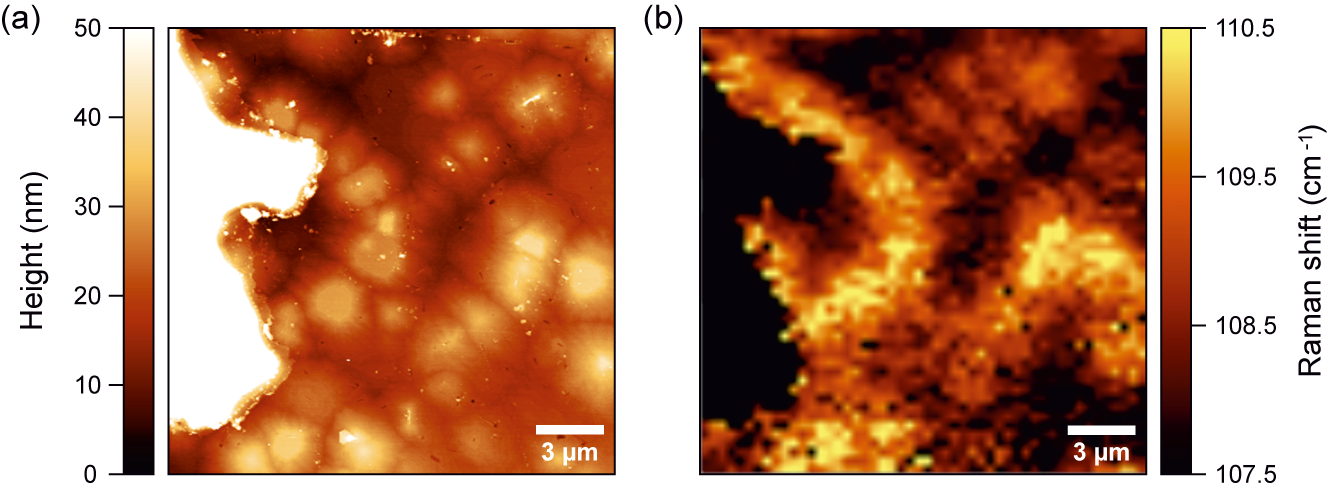}
	\caption{(a) Scanning force microscopy image of a (Bi$_{1-x}$Sb$_x$)$_2$Te$_3$ containing 63\% Sb. The elevated structure on the left hand side is a gold contact. (b) Corresponding Raman shift of the E$_g^2$ mode for the same region.}
	\label{Fig4-Raman-AFM}
\end{figure*}

When comparing the different distributions of peaks positions in Figs.~\ref{Fig3-Raman}(c) and (d) it can be seen that the distribution is narrow for low and for high Sb contents while it broadens for intermediate contents. To further investigate this observation we compare a spatially resolved Raman map and the corresponding surface topography measured by scanning force microscopy (SFM) of the film containing 63\% Sb which shows the broadest distribution. To identify the same position for both measurements we used the edge of a gold contact on top of the film as a reference. The SFM measurement of a $20 \times 20\,\mu$m$^2$ area is shown in Fig.~\ref{Fig4-Raman-AFM}(a). A part of the gold contact can be identified on the left hand side whereas the rest of the image shows the topography of the MBE grown film. It is observed that the surface exhibits distinct variations in height and the formation of islands which has been previously reported for MBE grown films.\citep{Krumrain11,Lanius} Next, we probed the same area by micro Raman spectroscopy. The resulting positions of the E$_g^2$ is shown in Fig.~\ref{Fig4-Raman-AFM}(b). The black area on the left hand side again shows the position of the thick gold contact where no (Bi$_{1-x}$Sb$_x$)$_2$Te$_3$ spectrum can be measured. We observe a strong spatial shift of the peak positions. A similar behavior can be observed for the A$_{1g}^2$ mode (not shown). When comparing the peak shift to the SFM image it becomes apparent that the E$_g^2$ peak shifts to higher wavenumbers for larger SFM heights and vice versa. 

In principle there are three explanations for such an correlation. 
The first and most straightforward one is that the peak position just depends on the height of the film. We therefore investigated films containing 63\% Sb having thicknesses of 8\,nm, 16\,nm, 27\,nm, and 56\,nm. The measured average peak positions are summarized in Tab.~\ref{tabThicknessDependence}. From these measurements we found no clear dependence of the E$_g^2$ peak position on the film thickness. Also for related systems no thickness dependence in the range of 5 to 80\,nm were observed.\citep{Shahil12, Shahil10, Dang10} The second possibility could be a local strain which changes the bond length and therefore the energy of the measured phonon mode within the islands. It has been shown that a reduction of bond length leads to a blue shift of the Raman modes for pure bismuth telluride.\citep{Kullmann84} That would mean that the bond length is reduced within the islands. However, since the quintuple layers are bonded by van der Waals forces it is expected that strain does not play a prominent role. The third and most likely explanation is a phase separation of Sb-rich and Sb-poor regions which relates to the islands observed in the SFM image. As a matter of fact, in the ARPES spectra the observed states were rather smeared and unsharp. Since a large area is illuminated, the signal will be averaged over different regions with slightly varying Sb content. From the broadening of the ARPES spectra shown in Fig.~\ref{Fig3-Raman} we estimated an Sb variation of 5\% at most. Using the FWHM of the distributions of the modes shown in Figs.~\ref{Fig3-Raman}(c) and (d), we find a variation of $\sim{8.5}$ percentage points from the E$_g^2$ mode and $\sim{4.8}$ percentage points from the A$_{1g}^2$ mode for the film containing 63\% Sb. These values are averaged over the spot size with a diameter of about 500\,nm and they agree well with the estimations from the ARPES spectra. Additional measurements like energy dispersive X-ray spectroscopy might be useful to support this explanation. 

The local variations in the measured Raman spectra can thus be used to gain information on the homogeneity of the samples. Interestingly, a comparison of the roughness of the film (Tab.~\ref{tab:table-1}) and the width of the distribution of the peak positions shows that flatter samples exhibit a more narrow distribution in the peak position than rougher ones. To confirm this statement we slightly adjusted the growth parameters (slightly slower growth rate and better base vacuum) which results in smoother films and considerably less variation in the peak position. These films (Sb content of 43\%, 45\%, 59\%, and 70\%) are added in Fig.~\ref{Fig3-Raman}(b) (solid symbols). Therefore, micro Raman measurements can be regarded as a versatile tool to analyze the growth quality of such films.
\begin{table}\centering
	\caption{Position of the E$_g^2$ and the A$_{1g}^2$ mode for different film thicknesses $d$. The Sb content is 63\% for all films.}
	\label{tabThicknessDependence}
	\begin{tabular*}{1.0\columnwidth}{@{\extracolsep{\fill}} c c c }
		\hline \hline \\[-7pt]
		$d$\,(nm)	& E$_g^2$\,(cm$^{-1}$)	& A$_{1g}^2$\,(cm$^{-1}$)	\\ \\[-8pt]
		\hline \\[-8pt]
		$8$			& $107.4 \pm 0.1$		& $155.2 \pm 0.2$			\\
		$16$		& $108.1 \pm 0.2$		& $155.5 \pm 0.3$			\\
		$27$		& $109.0 \pm 0.2$		& $156.6 \pm 0.3$			\\
		$56$		& $107.5 \pm 0.2$		& $153.2 \pm 0.6$			\\ \\[-8pt]
		\hline \hline
	\end{tabular*}
\end{table}

\section{Magnetotransport}

\begin{figure}[htb]
	\includegraphics[width=1.00\columnwidth]{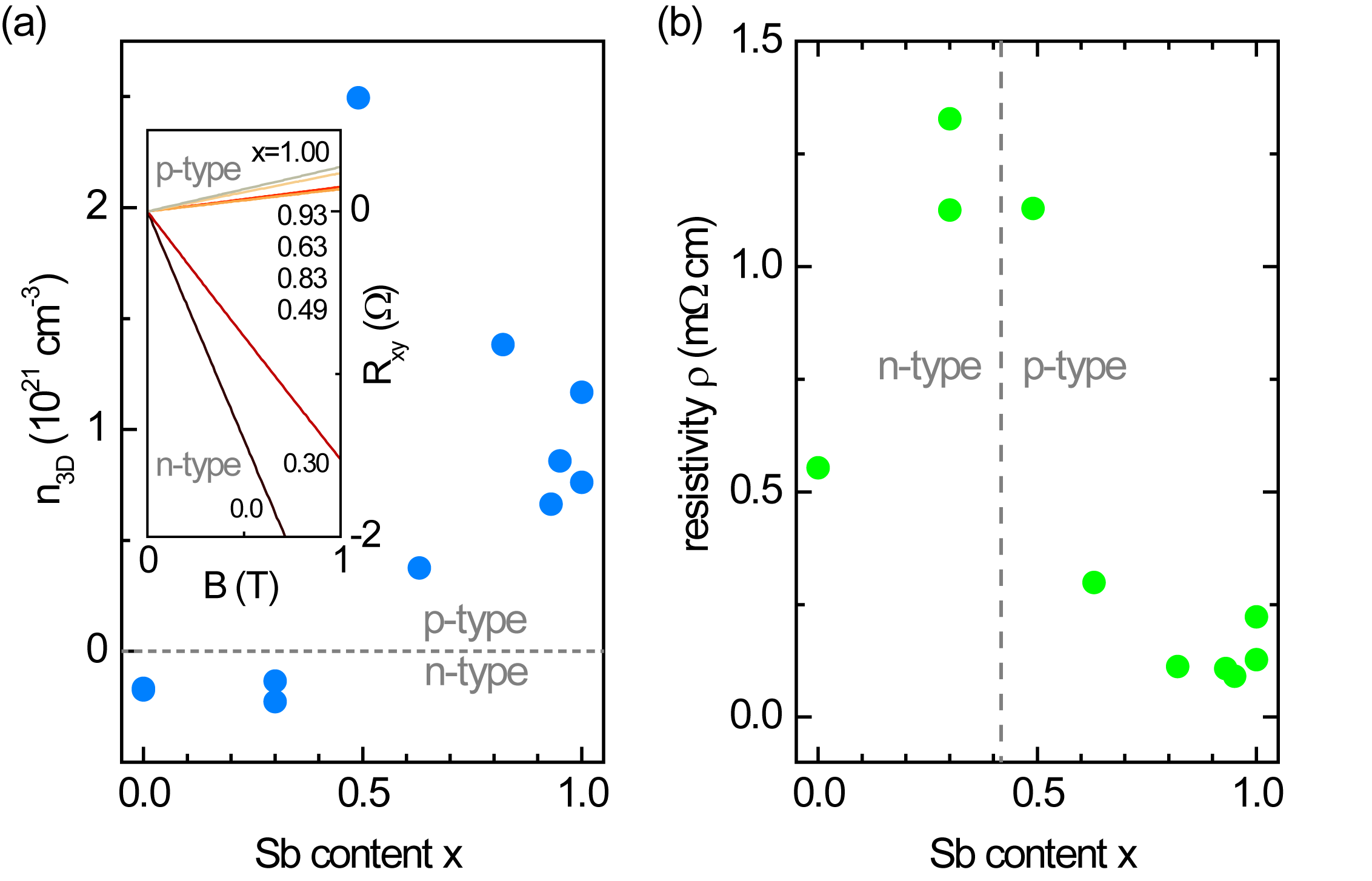}
	\caption{(a) Carrier concentration $n_\mathrm{3D}$ of (Bi$_{1-x}$Sb$_x$)$_2$Te$_3$ layers with different Sb contents $x$. The measurements were performed at a temperature of 1.8\,K. The inset shows the Hall resistance $R_\mathrm{xy}$ as a function of magnetic field $B$ for different Sb contents $x=100$\%, 93\%, 82\%, 63\%, 49\%, 30\%, and 0.0\%.  (b) Corresponding resistivity $\rho$ for different Sb contents.}
	\label{Fig5-BST300-n_3D-rho}
\end{figure}

In order to assess the transport properties of the (Bi$_{1-x}$Sb$_x$)$_2$Te$_3$ films, van der Pauw measurements were performed. In Fig.~\ref{Fig5-BST300-n_3D-rho}(a) the three-dimensional carrier concentration $n_\mathrm{3D}$ for layers with an Sb content between 0 and 100\% (samples A1 to A8) is plotted. 
The values of $n_\mathrm{3D}$ were extracted from the inverse slope of the Hall resistance $R_\mathrm{xy}$ as a function of magnetic field $B$. As can be seen in Fig.~\ref{Fig5-BST300-n_3D-rho}(a), inset, $R_\mathrm{xy}$ depends linearly on $B$. 
For the samples with $x < 49$\% the slope is negative, showing $n$-type, while for larger Sb contents the slope is positive, i.e. the transport is $p$-type. In the $p$-type range the slope of $R_\mathrm{xy}$ is relatively small indicating a large hole concentration. Generally, the hole concentrations are found to be roughly in the range between $3.7 \times 10^{20} $ and $1.4 \times 10^{21}$\,cm$^{-3}$. The exceptionally large value at $x=49$\% has to be taken with caution, since here transport in compensating $p$- and $n$-type channels probably leads to a decrease of the slope. The obvious discrepancy with the ARPES results on this particular sample can be explained by the fact that ARPES is a highly surface sensitive method, while electrical transport probes the entire system which can behave substantially different. As can be inferred from Fig.~\ref{Fig5-BST300-n_3D-rho}(a), compared to the $p$-type samples the carrier concentration of the $n$-type layers is systematically smaller. Here, electron concentrations in the range between $1.3 \times 10^{20}$ and $2.3 \times 10^{20}$\,cm$^{-3}$ are found.  However, because of the large errors connected to the contact placement of the van der Pauw samples, the values given above should be taken as a rough estimate only. For our set of samples the transition between $n$- and $p$-type transport is found at an Sb content between 30\% and 63\%, which is in the same range observed by \citet{He12} However, these values are considerably smaller than the transition at about 94\% reported by \citet{Zhang11} As can be seen in Fig.~\ref{Fig5-BST300-n_3D-rho}(b), the highest resistivity $\rho$ is measured in the transition range between $n$- and $p$-type conductance. The systematically lower values of $\rho$ for $x$ larger than 50\% is consistent with large carrier concentration for the $p$-type layers.   

\begin{figure}[htb]
	\includegraphics[width=1.00\columnwidth]{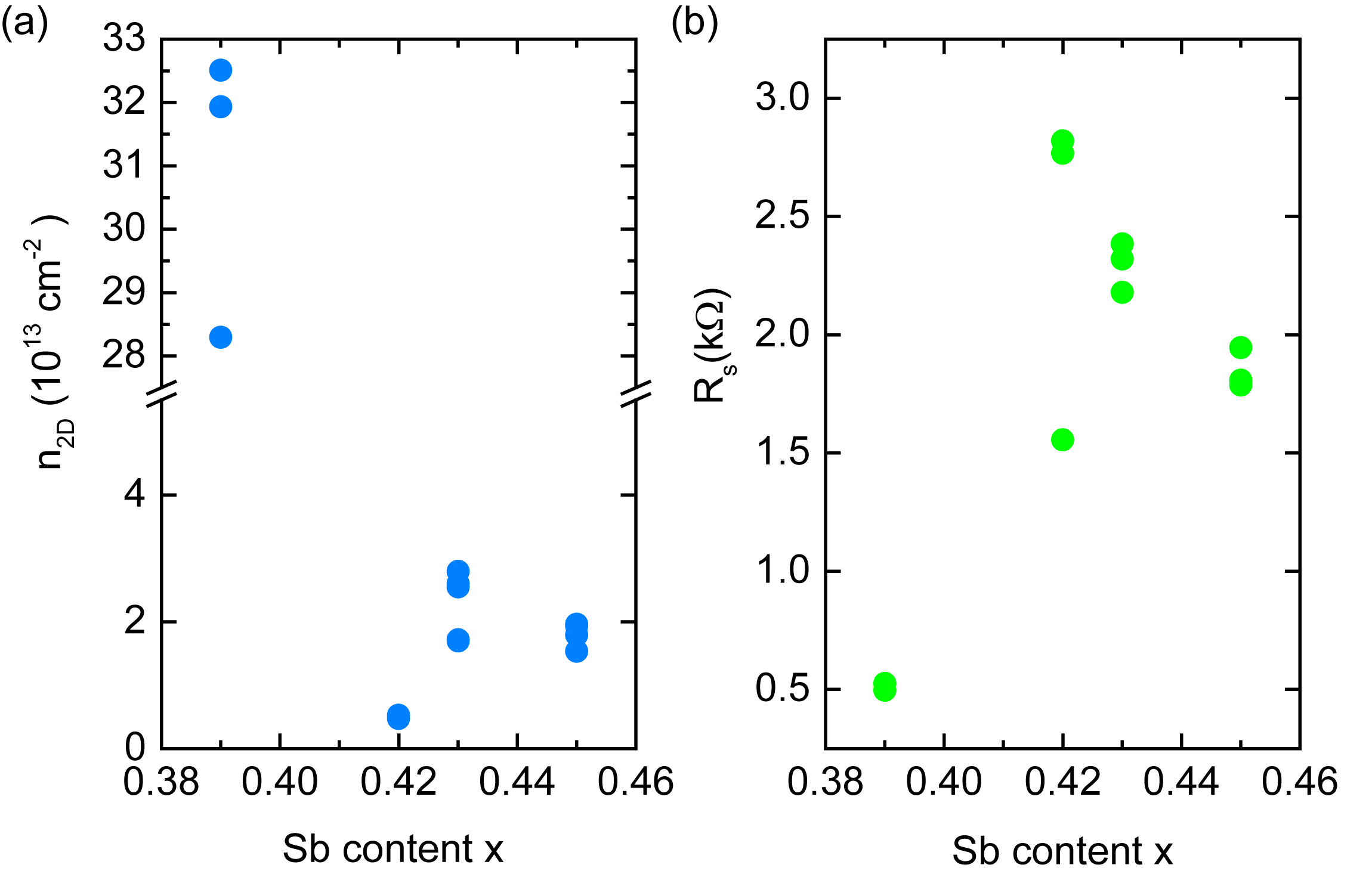}
	\caption{(a) Absolute values of the sheet electron concentration $n_\mathrm{2D}$ for (Bi$_{1-x}$Sb$_x$)$_2$Te$_3$ layers with an Sb content between 39\% and 45\%. (b) Corresponding sheet resistance $R_\mathrm{s}$.}
	\label{Fig6-BST323-326-n2D-Rsheet}
\end{figure}

\begin{figure}[htb]
	\includegraphics[width=0.95\columnwidth]{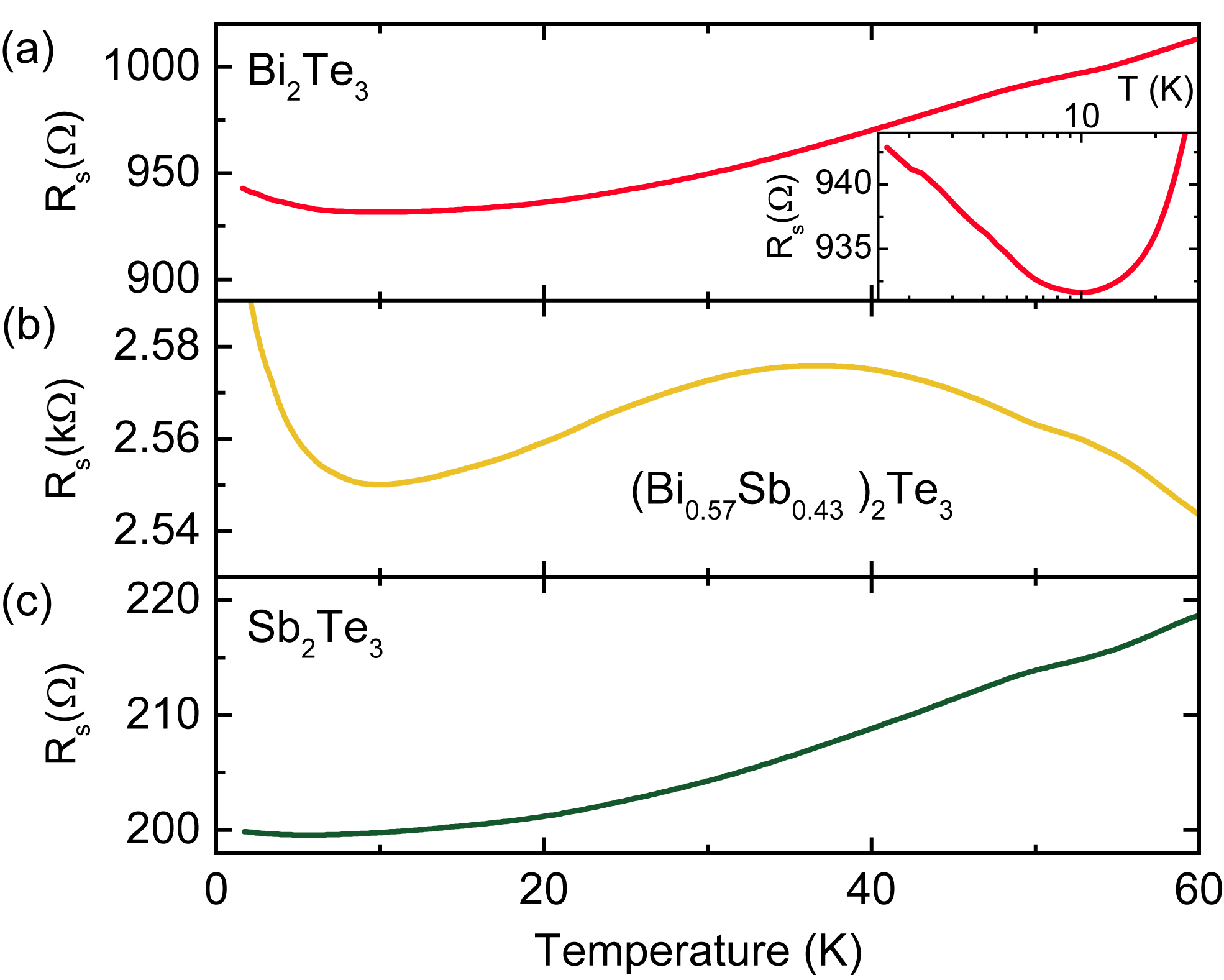}
	\caption{Sheet resistance $R_\mathrm{s}$ for (a) binary Bi$_2$Te$_3$, (b) the ternary layer with 43\% Sb content and (c) binary Sb$_2$Te$_3$. The insert in (a) shows a clear ln(T)-dependence of $R_\mathrm{s}$ in Bi$_2$Te$_3$ below 10K.}
	\label{Fig7-comparison_Rxx-T_Bi2Te3_Sb2Te3}
\end{figure}

\begin{figure}[htb]
	\includegraphics[width=0.95\columnwidth]{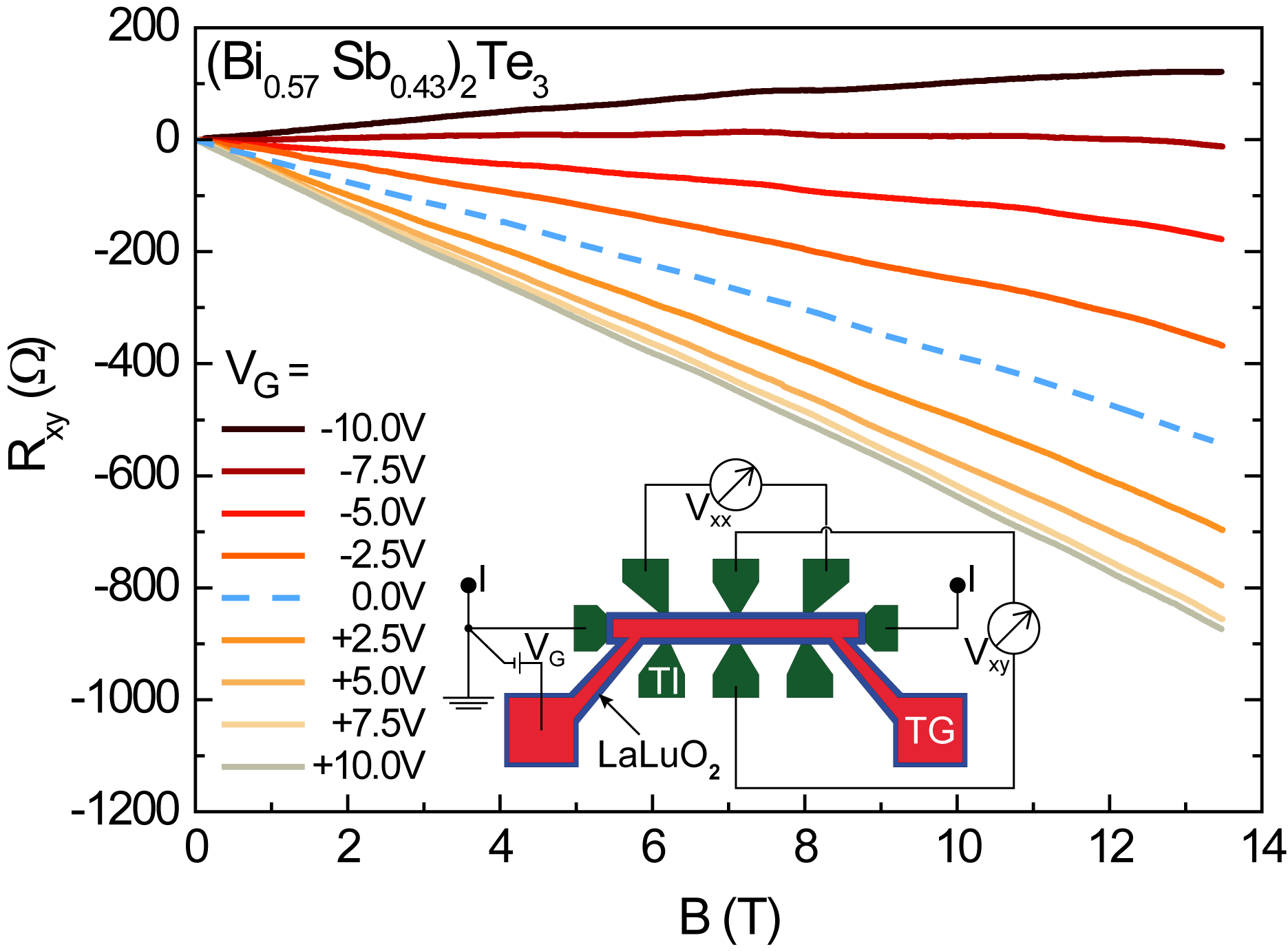}
	\caption{Hall resistance $R_\mathrm{xy}$ as a function of magnetic field of a (Bi$_{0.57}$Sb$_{0.43}$)$_2$Te$_3$ layer. The gate voltage was varied between $-10$ and $+10$\,V. The temperature was $1.8$\,K. A schematic depiction of the measurement setup is given by the insert.}
	\label{Fig8-BST326_Rxy-B-gate}
\end{figure}

\begin{figure}[htb]
	\includegraphics[width=1.00\columnwidth]{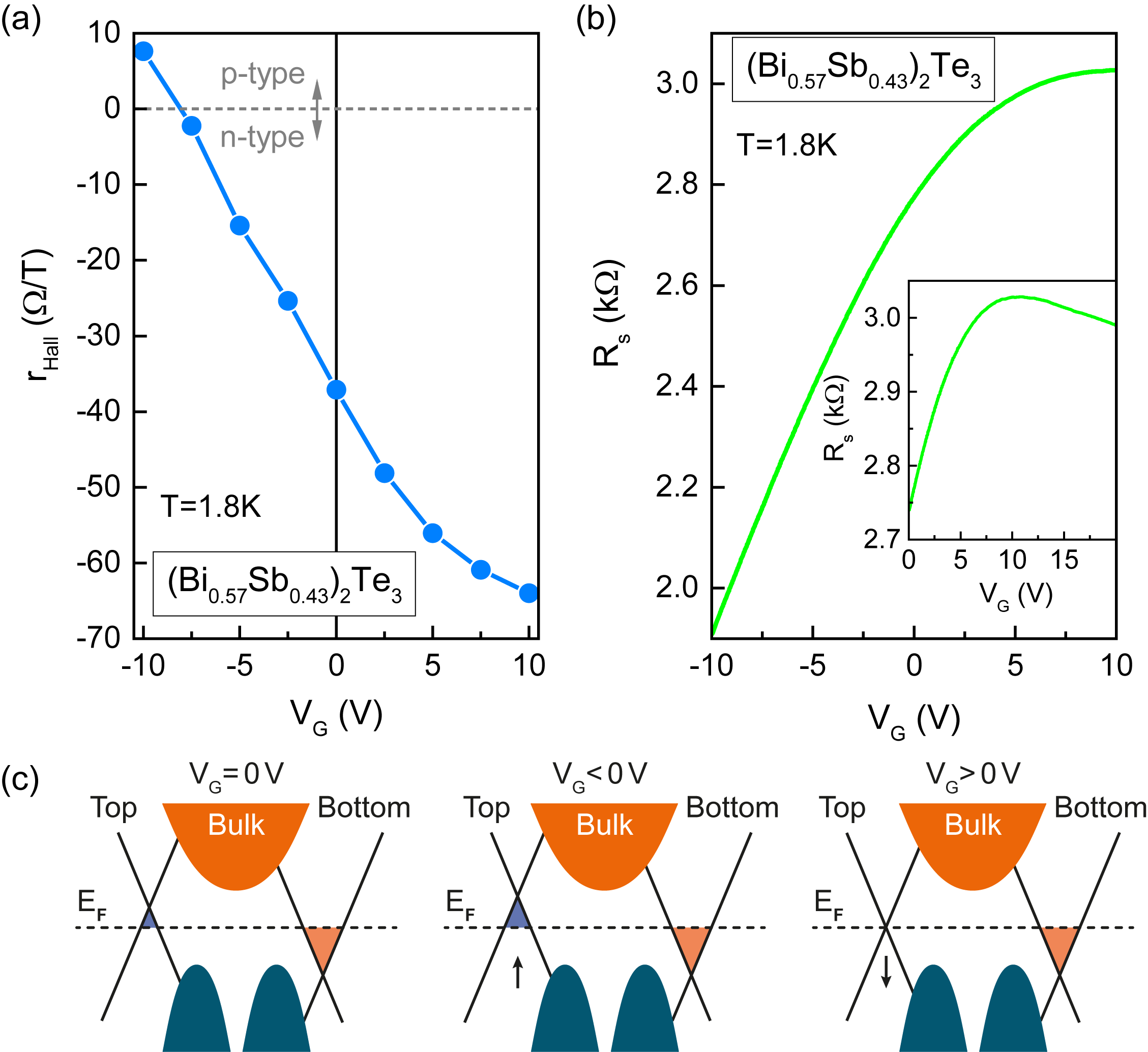}
	\caption{(a) Hall constant $r_\mathrm{Hall}$ of a (Bi$_{0.57}$Sb$_{0.43}$)$_2$Te$_3$ layer as a function of gate voltage $V_\mathrm{G}$. The measurements were performed at a temperature of 1.8\,K. (b) Corresponding measurement of the sheet resistance $R_\mathrm{s}$ as a function of $V_\mathrm{G}$. The inset shows a measurement of $R_\mathrm{s}$ for gate voltages from 0 up to 20\,V. (c) Schematic evolution of the top and bottom surface states when applying a gate voltage $V_\mathrm{G}$.}
	\label{Fig9-BST326_E4_60-rHall-Rs-gate}
\end{figure}

Based on the results given above, a new set of samples (B1 to B4) was prepared with an Sb content between 39\% and 45\% being close to the previously observed transition between $n$- and $p$-type conductance. In order to allow for a more precise determination of the transport parameters, geometrically well-defined Hall bars were prepared with dimensions as stated in Sec. II and measured. From now on, we can suspect that transport may no longer be completely dominated by bulk conductance, hence it becomes feasible to refer the data to two-dimensional systems. In Fig.~\ref{Fig6-BST323-326-n2D-Rsheet} the sheet carrier concentration $n_\mathrm{2D}$ and sheet resistance $R_\mathrm{s}$ are given for Sb contents between 39\% and 45\%.
All samples showed $n$-type conductance, no transition to $p$-type transport was observed for increasing Sb content. For an Sb content of $39$\% still relatively large values of $n_\mathrm{2D}$ and corresponding lower values of $R_\mathrm{s}$ are found. The lowest value for $n_\mathrm{2D}$ and the highest sheet resistance is found for $x=42$\%. In fact, the sheet electron concentration of $n_\mathrm{2D}=5 \times 10^{12}$ cm$^{-2}$ is in the range expected for topologically protected surface states. The low carrier concentration indicates that the Fermi level is located within the bulk band gap, i.e. the carriers are mainly provided by the topologically protected surface states. For larger Sb contents of 43\% and 45\% $n_\mathrm{2D}$ is slightly larger but does not exceed $3 \times 10^{13}$ cm$^{-2}$. This is more than what can be expected to originate from the surface states alone (about $1 \times 10^{13}$ cm$^{-2}$ per surface state with a band gap around 150\,meV) but the bulk transport should still be severely diminished. This statement is also supported by the temperature dependence of the sheet resistance. As can be seen in Fig.~\ref{Fig7-comparison_Rxx-T_Bi2Te3_Sb2Te3}(b), for the sample with an Sb content of $43$\%, an insulating behavior is found for temperatures above 40\,K, i.e. decreasing $R_\mathrm{s}$ with increasing temperature. Below 40\,K  $R_\mathrm{s}$ decreases with decreasing temperatures, while it once again increases below 10\,K. A very similar behavior was found by Zhang \emph{et al.} \citep{Zhang11} as well as by Shimizu \emph{et al.} \citep{Shimizu12}, although at larger Sb contents. Here, the temperature dependence above 40\,K was attributed to the Fermi level being located within the bulk band gap, so that carriers are thermally excited into the conduction band. A similar behavior has also been reported for Bi$_2$Se$_3$ after doping it with Pb to reduce the intrinsic carrier concentration.\citep{Wang2011} We refrain from further quantitative analysis of any parameters like activation energy or localization length if one were to consider variable range hopping transport, since the data available is insufficient to draw definite conclusions. For comparison, the temperature dependence of the sheet resistance for binary materials Bi$_2$Te$_3$ and Sb$_2$Te$_3$ is shown in Fig.~\ref{Fig7-comparison_Rxx-T_Bi2Te3_Sb2Te3}(a) and (c). They exhibit metallic behavior as $R_\mathrm{s}$ increases with temperature.  This clearly indicates that in these compounds, the Fermi level is located within in bulk conduction (Bi$_2$Te$_3$) and valence band (Sb$_2$Te$_3$), respectively. The upturn in resistance for temperatures below 10\,K (which is more pronounced for Bi$_2$Te$_3$) is often attributed to electron-electron interaction\citep{Wang2011,Zhao2014} which gives rise to a logarithmic temperature dependence for two-dimensional systems. This can be seen in the insert of Fig.~\ref{Fig7-comparison_Rxx-T_Bi2Te3_Sb2Te3}(a).
For the Hall bar sample with an Sb content of 43\% we were able to control the carrier concentration by applying a voltage $V_\mathrm{G}$ to a top gate electrode. As can be seen in Fig.~\ref{Fig8-BST326_Rxy-B-gate}, initially at $V_\mathrm{G}=0$ the transport is $n$-type, indicated by the negative slope of the Hall resistance $R_\mathrm{xy}$ vs. $B$ curve. 
By applying a positive gate voltage, the negative slope increases. However, with increasing negative gate bias, a transition from $n$- to $p$-type transport is observed. This is also illustrated in Fig.~\ref{Fig9-BST326_E4_60-rHall-Rs-gate}(a), where a monotonic decrease of the Hall constant $r_\mathrm{Hall}=\Delta R_\mathrm{xy}/\Delta B$ is seen with $V_\mathrm{G}$, with a change of sign at around $V_\mathrm{G}=-7.5$\,V.          
This is accompanied by a monotonous increase of the sheet resistance $R_\mathrm{s}$ for increasing $V_\mathrm{G}$, with a tendency to saturate towards $V_\mathrm{G}=+10$\,V. 

The continuous transition of $r_\mathrm{Hall}$ between $p$- and $n$-type is inconsistent in a single channel picture. In fact, here $r_\mathrm{Hall}$ should diverge at the transition point between $n$- and $p$-type conductance because of the diminishing carrier concentration and the corresponding inverse dependence of $r_\mathrm{Hall}$. Although this could also arise from percolations in an inhomogeneous system (due to Sb- and Bi-rich islands, as stated before), it is quite unlikely the case for this sample. While the difference in Sb concentration was found to be about 5\% for the worst case sample ($x=63$\%), it is most likely to be less in this sample which is much smoother (see Table~\ref{tab:table-1}). This has already been pointed out earlier to lead to narrower peak distributions in the Raman spectrum. Therefore, the gradient in local carrier concentration should not be sufficient to produce signatures of multi channel transport. Another possible scenario that needs to be discussed arises from the high degree of electrical compensation in the bulk which could cause the sample to act like a dirty metal. These systems are also known to exhibit nontrivial behavior of the longitudinal and transversal resistances and the observed temperature dependence of $R_{s}$ as well as the linear magnetoresistance at higher fields do not refute that possibility. \cite{Parish2003} Such a system however, would host only one transport channel (the bulk) which is in disagreement with our results from analyzing the weak antilocalization feature as will be discussed later in this section. Instead we find a satisfying explanation when we assume a conductance through two independent parallel channels. This scenario is also supported by the fact that the Hall curves shown in Fig.~\ref{Fig8-BST326_Rxy-B-gate} are slightly nonlinear, which indicates the presence of two channels with different mobilities and carrier concentrations. From the sign change of the slope at $V_\mathrm{G}=-7.5$\,V it can even be deduced that there exists a $p$- and $n$-type channel in parallel. Since the non-linearity of $R_\mathrm{xy}$ with $B$ is relatively weak we refrained from extracting the transport parameters of each channel from a two-channel fit.\citep{Taskin12,Bansal12} However, some qualitative conclusions can be drawn. According to the ARPES measurements shown in Fig.~\ref{Fig2-ARPES} and to band structure calculations of \citet{Zhang09} for (Bi$_{0.57}$Sb$_{0.43}$)$_2$Te$_3$ the Dirac point of the topologically protected surface states is expected to lie below the valence band maximum, while the transition from $n$- to $p$-type occurs at roughly 90\% Sb concentration. In order to qualitatively explain the smooth sign change of $r_\mathrm{Hall}$ we refer to the scheme shown in Fig.~\ref{Fig9-BST326_E4_60-rHall-Rs-gate} (c). At zero gate voltage we assume the electrochemical potential to be located within the bulk band gap at both top and bottom surface of the film by looking at the thermally activated behavior of the sample shown in Fig.~\ref{Fig7-comparison_Rxx-T_Bi2Te3_Sb2Te3} (b). This is further supported by the measurements of $R_{s}$ shown in Fig.~\ref{Fig9-BST326_E4_60-rHall-Rs-gate}(b), inset, where a gate bias up to $+20$\,V was applied. Here, a maximum of $R_{s}$ is observed at a gate voltage around $+10$\,V, suggesting that the electrochemical potential is located within the bulk band gap, i.e. in a low density of states region. Therefore, the two independent transport channels mentioned above can be attributed to the protected surface states at the interfaces between the topological insulator and the high-$k$ dielectric at the top side of the layer as well as the Te-passivated Si-substrate on the bottom side. Since the environment differs for both surfaces, the position of the Dirac point relative to the Fermi energy $E_F$ can in general be different. We assume it to be above $E_F$ at the top and below $E_F$ on the bottom. At least for the top surface this scenario is closer to the results found by ARPES for 93\% Sb concentration, rather than 43\%. We attribute this discrepancy to the different Fermi level pinning when the layer is covered by a dielectric layer on the top while lying on a Te passivation layer at the bottom.\citep{Kampmeier} Furthermore, due to the high dielectric constant of these materials ($\epsilon \geq 50$)\citep{Madelung1998} the bottom surface is almost completely shielded from the gate effect, such that the Dirac cone here is not expected to change with $V_\mathrm{G}$. The overall $n$-type character of $r_\mathrm{Hall}$ at $V_\mathrm{G}=0$\,V is then attributed to the larger electron density of states in the bottom surface compared to the holes available on the top. Upon decreasing $V_\mathrm{G}$ towards negative gate voltages, the upper Dirac cone is shifted upwards until the hole channel reaches a similar density of states as the electron channel on the bottom at which point both will cancel out in the Hall signal. This would explain both the smooth transition of $r_\mathrm{Hall}$ as well as the decreasing resistance at negative gate voltages, since the total number of charge carriers simply keeps increasing. Whereas, for $V_\mathrm{G}$ biased towards positive voltages the Dirac cone is shifted downwards with regard to the electrochemical potential so that $E_F$ crosses the Dirac point, resulting in a resistance maximum as explained before without changing the $n$-type character of the Hall response. Another possible scenario to be considered is that electron-hole puddles can form at low carrier concentrations. When the potential landscape on the surface is not uniform, there might be areas of the sample in which the chemical potential lies above the Dirac point while simultaneously being located below it in others. This can also lead to a smooth transition of $r_\mathrm{Hall}$ from n- to p-type character and might artificially reduce and broaden the maximum resistance as well as shift the actual point of transition but it does not explain the discrepancy between the transition point being at $V_\mathrm{G}=-7.5$\,V and the resistance maximum at $+10$\,V. Thus, we cannot conclude that no such puddles do form on the surface, but it has no fundamental impact on the model described above.

\begin{figure}[htb]
	\includegraphics[width=0.99\columnwidth]{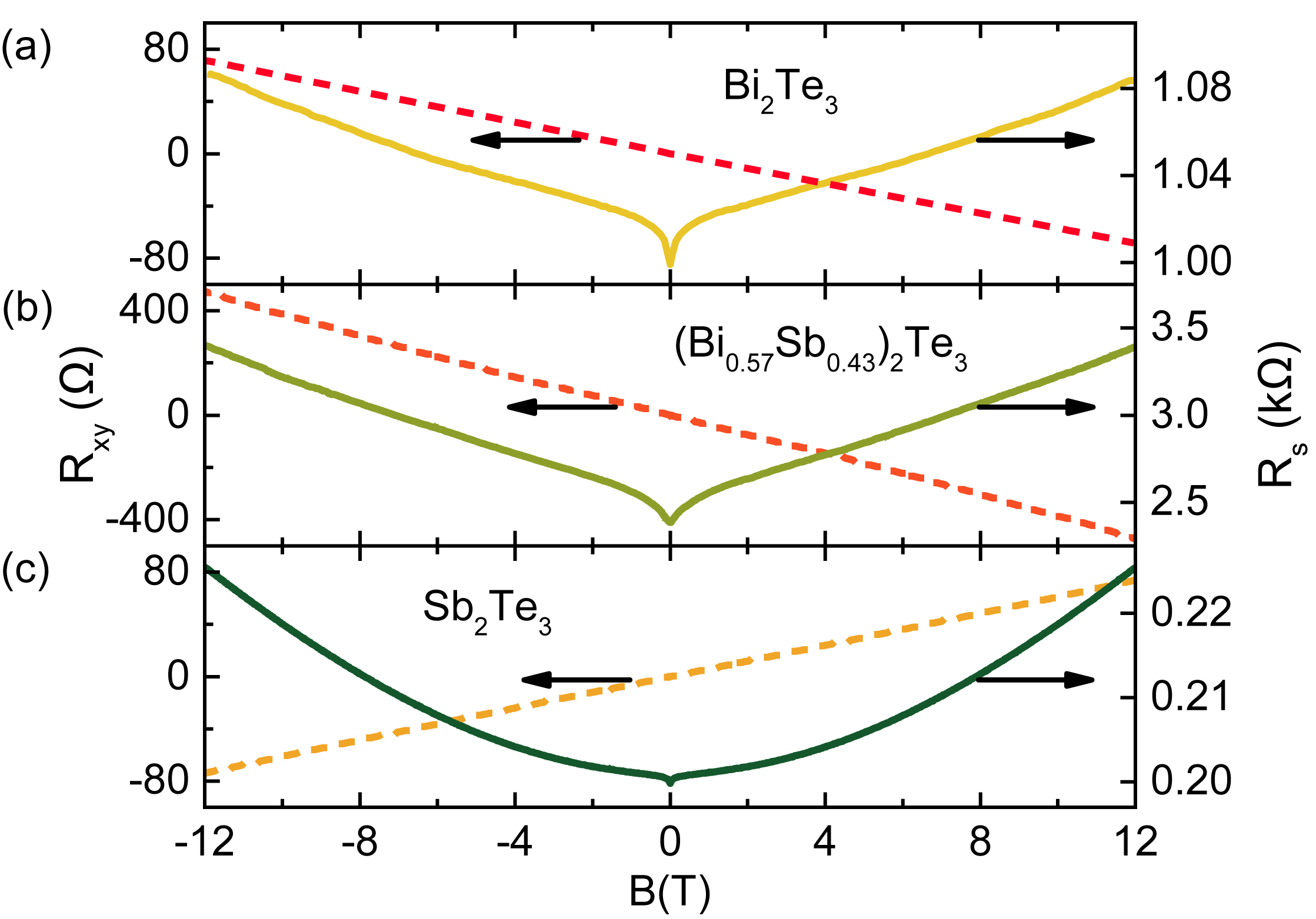}
	\caption{Hall resistance $R_\mathrm{xy}$ and sheet resistance $R_\mathrm{s}$ of a  (Bi$_{1-x}$Sb$_x$)$_2$Te$_3$ layer with different Sb contents $x$: (a) 0\%, (b) 43\%, and (c) 100\%, respectively. The measurements were performed at a temperature of 1.8\,K.}
	\label{Fig10-comparison_Rxx-Rxy_Bi2Te3_Sb2Te3}
\end{figure}

Further insight into the transport properties of topological insulators can be gained from the weak antilocalization effect in the magnetoresistance close to zero field.\citep{Zhang12b,Roy13,Tkachov13,Ockelmann2015} Weak antilocalization is observed in topological insulators, since spin-momentum locking leads to a suppression of time-reversal coherent backscattering. As a result the conductance is enhanced at zero field. This effect can be seen in Fig.~\ref{Fig10-comparison_Rxx-Rxy_Bi2Te3_Sb2Te3}, where $R_\mathrm{s}$ is plotted as a function of $B$ for Bi$_2$Te$_3$, (Bi$_{0.57}$Sb$_{0.43}$)$_2$Te$_3$, and Sb$_2$Te$_3$, respectively. 
As can be inferred from the corresponding Hall measurements also shown in Fig.~\ref{Fig10-comparison_Rxx-Rxy_Bi2Te3_Sb2Te3}, for the first two samples the transport is $n$-type, while it is $p$-type for Sb$_2$Te$_3$. For the ternary material the sheet resistance is considerably higher and the slope of the  Hall resistance is larger indicating a reduced carrier concentration compared to the binary samples. This once again confirms the compensation discussed above. The transport data of these samples are summarized in Table~\ref{tab:table-3}. 
\begin{table}\centering
	\caption{Parameters of the samples analyzed by means of the weak antilocalization effect: sample, sheet resistance $R_{s}$, mobility $\mu$, carrier concentration, pre-factor $\alpha$, and phase coherence length $l_\phi$.  \label{tab:table-3}}
	\begin{tabular}{cccccc}
		\hline \hline \\[-7pt]
		Sample 	& $R_{s}$ & $\mu$ & $n$, $p$ & $\alpha$ & $l_\phi$ \\ 
				& (k$\Omega$) & (cm$^2$/Vs) & ($10^{19}$ cm$^{-3}$) & & (nm) \\ \\[-8pt]
		\hline \\[-8pt]
		Bi$_2$Te$_3$ & 1 & 50 & $8$ & $-0.5$ & 210 \\
		(Bi$_{0.53}$Sb$_{0.47}$)$_2$Te$_3$ & 2.5 & 150 & $0.6$ & $-1.37$ & 72 \\
		Sb$_2$Te$_3$ & 0.2 & 300 & $5$ & $-0.5$ & 430 \\ \\[-8pt]
		\hline \hline
	\end{tabular}
\end{table}

A detailed analysis of the weak antilocalization effect can be carried out by fitting the experimental data to the Hikami--Larkin--Nagaoka model \citep{Hikami80}
\begin{equation}
\Delta \sigma (B)=-\alpha \frac{e^2}{2\pi^2 \hbar} \left[\ln
\left(\frac{B_\phi}{B}\right) - \psi
\left(\frac{B_\phi}{B}+\frac{1}{2}\right) \right]\; . \label{Eq:sigmaperp} 
\end{equation}
Here, $\psi$ is the digamma-function and $B_\phi$ is the characteristic field defined by $B_\phi=\hbar/(4|e|l_\phi^2)$. The latter quantity contains the phase coherence length $l_\phi$, which is a measure for the distance over which the carriers can propagate phase-coherently. For single transport channels with strong spin-orbit coupling the pre-factor $\alpha$ in Eq.~(\ref{Eq:sigmaperp}) is expected to be $-1/2$. In case of more than one transport channel $\alpha$ will deviate from this value.

In Table~\ref{tab:table-3} the values of $\alpha$ and $l_\phi$ for the three measured samples are given. For the binary materials $\alpha$ has a value of $-0.5$ indicating that only a single transport channel is present. This is also in good agreement with previous studies on these materials (see Refs. [\citenum{He2012,Liu2011,Takagaki12,He2011,Wang2011}] and references therein). Whereas for the ternary layer a value of $-1.37$ is extracted, which is close to $-1.5$ expected for three transport channel. Thus in this case it might be possible that the weak antilocalization effect originates from the contribution of two surface channels, i.e. at the upper surface, at the interface to the substrate, and in the bulk channel although its contribution is strongly suppressed compared to the binary systems. This finding also is not in favor of the film just acting as a dirty metal as stated earlier and giving rise to the observed transition in the Hall signal as a pure bulk effect. It furthermore deviates from findings on other composite systems like Bi$_2$Se$_3$/In$_2$Se$_3$ superlattices\citep{Zhao2014} or Bi$_2$Se$_3$/Bi$_2$Te$_3$ heterostructures\citep{Zhao2013} in which the results can be explained by a stack of individual TI layers or show no signs of multiple decoupled transport channels within a single film. The fact that for the binary materials the pre-factor is $-0.5$ indicates the presence of only a single channel. This can be attributed to a domination of the bulk contribution to the transport, which is also supported by the observation that the sheet carrier concentration is much higher than for the ternary material. For Sb$_2$Te$_3$ a phase coherence length exceeding 400\,nm is extracted. One finds that $l_\phi$ of the binary layers is considerably larger than for the ternary material. The underlying mechanism is not clear yet. It might be due to enhanced disorder or due to the fact that for the ternary material, the surface states contribute to the weak antilocalization effect, which is masked by the bulk contribution for the binary layers. A higher $l_\phi$ would also allow for more opportunities for carriers to scatter coherently between surface and bulk states, leading to a stronger coupling which would make them appear as one effective channel. This coupling could be reduced in the ternary system even if the bulk has some residual contribution to transport.

\section{Conclusion}

Combining a multitude of techniques, we were able to access a broad spectrum of properties of (Bi$_{1-x}$Sb$_x$)$_2$Te$_3$, including growth and film quality, electronic structure as well as transport properties. By using MBE a set of (Bi$_{1-x}$Sb$_x$)$_2$Te$_3$ layers with an Sb content in the full range between 0 and 100\% were grown. Regarding the determination of the Sb content consistent results were obtained using XPS and Raman spectroscopy. Additionally we could show that also the intensity ratio $I_{009}/I_{0015}$ in XRD can be employed to gain information on the Sb content. Here, the dependence between the $I_{009}/I_{0015}$ ratio and the Sb content was modeled by a non-linear relation. Using scanning micro Raman spectroscopy local variations of the E$^2_g$ and A$^2_{1g}$ peaks were found. The most likely explanation for this phenomenon are spatial variations in the Sb content. This conclusion is also supported by the ARPES spectra, where the observed spectral features were rather broadened. Therefore, one has to be cautious with respect to inhomogeneity of the composition when interpreting data that is spatially averaged (e.g. ARPES and transport). Furthermore, from the ARPES measurements we also found that for an Sb content of 49\%, the Dirac point lies well below the Fermi level. The Fermi level is located close to the bottom of the conduction band. Consequently, $n$-type conductance was observed in transport measurements. The situation changes for larger Sb contents. Here, the ARPES measurements revealed that the Fermi level shifts across the Dirac point. For Sb contents close to 100\% one can assume that at least for higher wave vectors the Fermi level crosses the valence band. For this case, the transport data show $p$-type conductance with a large bulk contribution. The trend observed in ARPES measurements is confirmed by magnetotransport measurements. With increasing Sb content a transition from $n$-type to $p$-type transport is observed. Moreover, in the transition range an increase of the resistivity was found, indicating that here the Fermi level is located within or close to the bulk band gap. On a layer with $x\left(\mathrm{Sb} \right)=43 $\% gate control of the charge carriers was achieved. The smooth transition of the Hall coefficient between $n$- and $p$-type upon gate biasing as well as the observed weak antilocalization indicated strongly that the transport takes place in two parallel channels with opposite character, i.e. $n$- and $p$-type. Still, more experiments are needed to completely rule out the possibility of the bulk contributing as a dirty metal in these films. To summarize, we can conclude, that by alloying two topological insulator materials with opposite type of conductance the Fermi level can be varied to a large extent, so that it can be shifted into the band gap region. Depending on the substrates that were used, previous studies find the transition from $n$- to $p$-type transport at Sb-contents that are either much higher (InP(111))\citep{Shimizu12} or very close (SrTiO$_3$(111))\citep{He12} to those in this work. This indicates that further studies on the relation between the substrate and the intrinsic carrier concentration of the films are needed. By choosing a different substrate it might also be possible to improve the crystal quality of the film and enhance the mobility.

\begin{acknowledgments}
The authors are grateful to Ewelina Hankiewicz and Grigory Tkachov (W\"urzburg University) for fruitful discussions, and H. Kertz and Ch. Krause for technical assistance. This work was pursued within the Virtual Institute for Topological Insulators (J\"ulich-Aachen-W\"urzburg-Shanghai), which is financially supported by the Helmholtz Association. B.B. and C.S. acknowledge support from DFG.   

\end{acknowledgments}

\bibliography{BiSbTe-ternary}

\end{document}